# E-Crime Legal Brief: A Case Study on Talk Talk Hacking

BONAVENTURE NGALA u1637893@uel.ac.uk

**Abstract**

E-crime has had various definitions for different countries and organizations. There is no universal definition of E-crime and therefore the interpretation is left to cybercrime investigators and judges to apply related crimes to within the scope where possible.

E-crime legal brief should include Citation, facts of the case, issues, reasoning, decision of the judges and analysis. The analysis outlines applicable laws in e-crime, case laws relevant to the facts of the case and crime committed. The analysis either provide supporting or contrary opinions diverting from ruling and decision made by the judges.

### 1.0 Case overview and definition of e-crime

It is an e-crime appeal case that involves hacking of TalkTalk Company leading to disclose of personal data, repeated attacks, online trade of obtained data, latency of its website and blackmail of the then Chief Executive Officer (CEO). Talktalk is a leading company that provides a range of broadband, telecommunication and network solutions services to various businesses across the UK.

E-crime in most jurisdictions has been interchangeably equated to cybercrime and has had various definitions depending on different jurisdiction and investigation institutions. It's for the benefit of this case that I explore various e-crime definitions and provide a definition that will be applied in the brief.

- **1.1.1** The Association of Chief Police Officers under e-crime strategy 2009 manual (ACPO, 2009) defines e-crime as the use of networked computer or internet technology to commit or facilitate the commission of crime.
- **1.1.2** Under its pronged approach to combat organized crimes, (United Nations, 1994) in its manual on prevention and control of computer related crimes defined e-crime to include fraud, forgery and unauthorized access.

**1.1.3** The UK parliament as per (Home affairs, 2013/2014) divides e-crime to: pure online crime where a digital system is the target as well as the means of attack, which include information technology infrastructure disruptions, data theft over network malware and includes also existing crimes that have transformed in scale or form by the use of the internet; (Stewart, 2006) or the use of internet to facilitate drug dealing, people smuggling and other traditional types of crimes.

**1.1.4** The legal dictionary (Stewart, 2006) refer to e-crime as a convenient name to describe some new crimes made possible by wider use availability of computer and the opportunities provided by the internet as well as some new ways of committing old crimes.

**1.1.5** (Begovic, 2017) defines e-crime as any criminal activity that involves the internet, computer or other device such as distribution in of distribution of electronic viruses, launching denial of service attack, fraud, harassment, copyright breaches and making, possessing or distributing objectionable material.

**1.1.6** According to (Yar, 2012) e-crime has been described the new patterns of vulnerability to criminal predation and harm resulting from transformation of mediated communication enabled by changing technologies.

**1.1.7** According (Gercke, 2012) describes e-crime as computer related crime that covers offences that are either related to a network or affect only stand-alone computer systems.

**1.1.8** E-crime (Malakedsuwan, 2019) has also been defined in the context of e-fraud as a crime that affects society, impacting upon individuals, businesses and governments.

**1.1.9** (Stefan Fafinski, 2010) even though alludes that there is no legal definition, he refers to e-crime as a crime committed by means of or with the assistance of the use of electronic networks.

**1.1.10** E-crime according to (Computer Hope, 2019) is an act performed by a knowledgeable computer user who illegally browses or steals a company or individual information.

**1.1.11** E-crime in this legal brief will be referred to as any crime committed or aided by use of electronic operations targeting computer systems security, data processed by them, disruptions of IT infrastructure, possession/distributing information by means of a computer system or network.

**1.2** Computer hackers (Computer Hope, 2019) possess special skills in coding computer programming tools or exploit vulnerabilities in a specific target device or network for access of secured information leading to either altering or disrupting system and or security features of either the device or network.

## 2.0 Case citation

Re V. Connor Douglas Allsop

2018/04878/A2

United Kingdom, Court of Appeal Criminal Division

30th January 2019

[2019] EWCA Crim 95; [2019]1 WLUK 434

Before: Lord Justice Flaux Mr Justice Sweeney and Mr Justice Soule

## 3.0 Facts of the case

**3.1** It's an appeal case for the sentence passed by a single Judge in a case that involved hacking into computer systems of a mobile phone communication company TalkTalk in October 2015.

**3.2** The appellant (Connor Douglas Allsop), 22 years of age at sentencing pleaded guilty on re-arraignment to count 6 "Supplying an article for use in fraud, contrary to section 7(1) of the Act (Fraud Act, 2006) and count 8 Supplying articles for the use in an offence under sections 1, 3 or 3ZA of the Act (Computer Misuse Act, 1990)"

**3.3** Mathew Hanley, co-accused pleaded guilty to count 3 "Causing a computer to perform a function to secure or to enable unauthorized access to a program or data, contrary to section 1(1) of the Act (Computer Misuse Act, 1990) count 4 Supplying articles for use in an offence under sections 1, 3, or 3ZA of the Act (Computer Misuse Act, 1990) count 5 supplying an article for use in fraud section 7(1) of the Act (Fraud Act, 2006) count 9 Obtaining articles for use in an offence under sections 1,3 or 3ZA of the Act (Computer Misuse Act, 1990)"

3.4 The court initially had decided the two defendants should wait trial outcome of Daniel Kelly also charged with number of cases relating to TalkTalk hacking attack and blackmail, but after several adjournment of the case, (Westlaw , 2019) on 19th November 2018 at the Central Criminal Court the appellant was sentenced to twelve months imprisonment on the two counts i.e. count 6 (eight months) and count 8 (concurrent term of seven months imprisonment).

3.5 BAE Systems was tasked to investigate the case after TalkTalk became aware of potential latency issues on its website as per (Westlaw , 2019), continued/repeated blackmail directed to then Chief Executive Officer (CEO) Dido Harding for demands of bitcoins in exchange of data stolen, National Criminal Agency (NCA) were informed of the attack.

3.6 The appellant and Hanley lived in Tamworth and knew each other, Hanley was a dedicated Computer hacker who according to (Westlaw , 2019) when he realized police were after him wiped his computer but fortunately police were able to obtain evidential information of his Skype conversation with the appellant detailing/admitting how he hacked to TalkTalk database and file exchange between the two.

3.7 The appellant (Westlaw , 2019) supplied the dump file to online user referred to as 'Reign' with full knowledge that Hanley and Reign were involved with fraud and hacking. The dump file consisted of personal and financial details of 8,000 TalkTalk customers. Reign later supplied the appellant with list of personal data and website details and passwords which he submitted to Hanley. The files were later used for further hacking.

3.8 The appellant in is plea (Westlaw , 2019) admitted having not known the impact of his actions to the victims while Hanley said his actions were not money motivated and bemoaned to have done such an activity.

3.9 Examination reports of the defendants brought to the Judges (Westlaw , 2019) assessed the appellant to be of low risk of offending and low risk of harm while Hanley was assessed to be of low risk to harm to others with probation officer considered him to be a vulnerable within a custodial environment due to his history of anxiety and low self-esteem therefore non-custodial sentence was preferred.

3.10 The judges in their sentencing remarks (Westlaw , 2019) noted that, both defendants were significantly involved in planning attacks of computer systems of TalkTalk Company, they also didn't report the vulnerabilities in the company systems, further, the actions of the two lead along with others to gain access to the company confidential data that included customers information, causing loss of £77 million, supplied confidential information to others causing misery and distress to thousands.

## 4.0 Issues

4.1 The appellant (Westlaw , 2019) raised three grounds of appeal in relation to the sentence given by the lower court judge as shown below.

    4.1.1 First, (Westlaw , 2019) did the judge overstate the appellant's involvement in taking part in the original hacking of TalkTalk database hence denying the appellant a lesser imprisonment by stating that the appellant was involved in a *"significant sophisticated, systematic, planned attack"*?

    4.1.2 Second, (Westlaw , 2019) did the judge distinguish sufficiently between the culpability of the two co-defendants, whereby unlike Hanley the appellant did not benefit from the attack hence raising the issue of culpability? *The appellant submitted that his culpability was less than that of Hanley and that the one-third less starting point of 20 months, rather than 30 months in case of Hanley, did not reflect their respective culpability, capitulations which the appellant cannot agree by.*

    4.1.3 Third, (Westlaw , 2019) did the judge err in not suspending the sentence? The appellant submitted to the court that, "*it should consider the period of delay between plea of guilty and sentence as the prospect of a custodial sentence was hanging over the appellant, that since sentencing the appellant has been incarcerated in Belmarsh prison for 20 weeks a sufficient sentence to reflect the seriousness of the offending.*"

## 5.0 Decision

5.1 The court of appeal dismissed the plea but agreed to correct a technical defect in the sentencing issued by the lower court. While delivering their ruling on (R v Connor Douglas Allsopp, 2019) the Judge's unanimously agreed that, "*the only amendment we would make to the sentences passed is that they should have been sentences of detention in a young offender institution, rather than imprisonment, because at the date of his conviction on 30th March 2017 the appellant was under 21 years.*"

## 6.0 Reasoning

6.1 **On the first issue**, (Westlaw , 2019) the court of appeal held that, "*we agree that if the judge's statement at the outset of her sentencing remarks that the appellant was involved in a **significant, sophisticated, systematic, planned attack** were taken in isolation, it might suggest that she had proceeded on the assumption that the appellant was involved to a greater extent than he was.*" However, the appeal court asserted that careful consideration of facts taken into account is clear that there is no question of the judge having misunderstood the extent of the appellant's involvement.

6.2 **On the second issue**, (Westlaw , 2019) court of appeal agreed with the lower court and stated that, "*the judge was right to put the appellant's offending towards the bottom end of the medium culpability category in the guideline and her starting point of 20 months faithfully reflected that.*" The appeal court judge' also noted that the lower court took into considerations the appellants youth, immaturity and delay in sentencing by reducing starting point from 20 months to twelve months, giving appellant full one-third credit after pleading guilty until four months after plea and case hearing and four months before trial which shows he was more generous. Therefore, the appeal court held that, (Westlaw , 2019) the sentence passed cannot be described as excessive, or disparity between appellant and Hanley.

6.3 **The third issue**, (Westlaw , 2019) the appeal court stated that, "*the judge took account of the delay in reducing the amount of the starting point and the issue in relation to whether or not she should have suspended the sentence is not one which is related to the delay but is related to the seriousness of the offending.*" Further the judges observed that the appellant did not initiate the attack but took advantage and facilitated fraudulent acts which could have affected a large number of victims. The issue of the lower court judge applying immediate custodial sentence was therefore appropriate.

## 7.0 Opinion

7.1 The court of appeal Judges agreed to the views of the single lower court judge in his decision during the initial sentencing. While delivering the ruling the Judge's noted that, (Westlaw , 2019) the sentencing attributed to the inadequacy of sentencing guidelines for offences under Computer Misuse Act 1990, she however based her sentence to (R v Martin 2013) and (R v Mudd 2017) where she related the following;

    7.1.1 Referring to (R v Martin, 2013) case, *"The offences fell into the highest level of culpability, mainly because of the financial loss and destruction to private and business affairs."*

    7.1.2 In (R v Mudd, 2017) case, the judge noted that, *"The appellant in the case, who was aged 20, had no serious convictions and was diagnosed with Asperger's Syndrome. The offence took place at age 16 and there was substantial mitigation. The court reduced the sentencing from six years for an adult to 24 months, the appeal court further reduced it to 21 months, but concluded that immediate custody was appropriate and said that it was important the courts sent clear message that cybercrime on this scale was not a game but would be taken very seriously by the courts and punished accordingly."*

7.2 On count 5 and 6 the judge referred the offences to section 7 of the (Fraud Act, 2006), where he found the offending on the lower end of medium culpability. *The judge noted that* (Westlaw , 2019)*, the appellant involvement was less than Hanley, she also considered pre-sentence report including submissions before the court. She noted that the appellant was 18 years at the time and immature, that he had no previous convictions, that the period between the offending and sentence was not appellants fault.*

> *7.2.1 This guided the judge in reducing the starting point from 20 months to 12 months; he further was given full one-third credit due to his considerable admissions in police interviews. The judge therefore in this view issued a concurrent imprisonment sentence as follows, 8 months sentence on count 6- and 7-months sentence on count 8.*
>
> *7.2.2 The judge also concluded that, she could not suspend the sentence in view of the seriousness of the offending.*

## 8.0 Analysis

8.1 The case is interesting and has significant impact to cases related to e-crime as it's among the cases that shocked the world on how young teenagers were able to compromise systems security of a big company.

8.2 The intentions of the hacker determine if the actions contain *mens rea* (Westlaw, 2019) if the hacker performs the following; decides to take the data to publish, blackmail legitimate owner with a threat of publishing the data, use the data to launch other attacks, duplicate the accessed database or does an illegal activity.

8.3 Although the judges were candid in delivering their opinion based on the case, but I found it lacking depth in terms of other case laws for informed decisions and wider scope of passing stiffer punishment as a way of deterrence.

8.4 Citing the definition of e-crime, facts of the case and submissions before the court, the following legislations were contravened; Computer Misuse Act 1990, The Frauds Act 2006, Protection to Harassment Act 1997, Criminal Damage Act 1971 and Data Protection Act 1998.

8.5 The act (Protection to Harassment Act, 1997) section 1 (1) a, b, (2) states that; *"(1) a person must not pursue a course of conduct, (a) which amounts to harassment of another, and (b) which he knows or ought to know amounts to harassment of the other, (2) for the purposes of this section, the person whose course of conduct is in question ought to know that it amounts to harassment of another if a reasonable person in possession of the same information would think the course of conduct amounted to harassment of the other."*

**8.6** Additionally, the act (Protection to Harassment Act, 1997) Section 8 (1) a, b as stated; "*every individual has a right to be free from harassment and, accordingly, a person must not pursue a course of conduct which amounts to harassment of another and, (a) is intended to amount to harassment of that person; or (b) occurs in circumstances where it would appear to a reasonable person that it would amount to harassment of that person.*"

**8.7** Therefore based on the understanding from (Protection to Harassment Act, 1997), that harassment amounts to any action taken by an individual *that may include alarming the person or causing the person distress* and based on the facts of the case it's clear that from his own admission the appellant had full knowledge of his actions which caused distress to TalkTalk CEO and customers as a result of repeated blackmail and therefore guilty of causing harassment.

**8.8** The reckless actions carried by both the co-accused also amounts to an offence under the act (Criminal Damage Act , 1971) section 1 (1); "*a person who without lawful excuse destroys or damages any property belonging to another intending to destroy or damage any such property or being reckless as to whether any such property would be destroyed or damaged shall be guilty of an offence.*"

**8.9** In the case of R v Whiteley 1991, (Casey, 2004) the court of appeal held that what the criminal damage act required to be proved was that tangible property had been damaged and not necessary that the damage itself should be tangible.

**8.10** While it may be challenging for the prosecution to proof in this case the accused acted intentionally, enough evidence based on admission interview exist to show that they acted recklessly. On his part the appellant offered to act as the link person between Hanley and online fraudsters with full knowledge the dump file contained customers personal details which were under custody of TalkTalk Company as one of their key assets for business operations. The company also reported potential website latency which was as a result of the hacking.

**8.11** Referring to the judge reasoning that the co-accused in (R v Connor Douglas Allsopp, 2019) "*was involved in a significant, sophisticated, systematic, planned attack,*" decision upheld by the appeal court, show the applicability of Data Protection Act 1998.

**8.12** The (Data Protection Act, 1998) Section 55 (1); "*A person must not knowingly or recklessly, without the consent of the data controller, (a) obtain or disclose personal data or the information contained in personal data, (b) procure the disclosure to another person of the information contained in personal data.*"

Therefore, careful analysis of the facts presented demonstrates clearly that the actions of the appellant and co-accused disclosed personal data and even went ahead to procure the disclosure of the data to online fraudsters leading to subsequent attacks, this therefore shows that the appellant committed an offence under Data Protection Act 1998.

**8.13** The issue of whether the evidence was enough raises the question of whether the court was right in sentencing the appellants based on the admission interview, data files and whatsapp conversation submitted in court. In the case of (Ellis v DPP, 2001), the judge while dismissing the appeal case on satisfactory of evidence noted that, "*the statutory provisions were sufficiently wide to include the use made of the computers by E. The evidence of the university administrative officer and the police was sufficient to justify a finding that there was a case to answer and the justices had been entitled to infer that E had the necessary intent and knowledge. It was not necessary for the prosecution to adduce expert evidence in a case alleging computer misuse; it was sufficient to rely on the admissions made in interview and the factual evidence of the administrative officer involved.*"

**8.14** The appeal claim case, (R v Connor Douglas Allsopp, 2019) the appellant pleaded that, although he knowingly participated in the "computer hacking" process by requesting for the dump file from Hanley and submitting it to online fraudster his actions were not severe and were driven by immaturity and desire to demonstrate dexterity to peers. The Judges were right to use the case of (R v Mudd, 2017) which demonstrates both considerate aspects of immaturity for young offenders whose desire to commit a crime is driven by the essence of showing their ability to peers, therefore taking into account the age of the offender while at the same time passing severe sentence is enough deterrence for any young cybercriminal with the same intentions or motives.

**8.15** Further analysis of (R v Mudd, 2017) case and that of (R v Connor Douglas Allsopp, 2019) shows sharp similarity in the aspects of public interest, parties involved, defendants actions among other aspects as below;

- **8.15.1** The cases were both appeal and were initially decided by a single judge at criminal magistrate court
- **8.15.2** Among the counts the most predominant were (supplying an article for use in an offence contrary to section 1 or 3 of (Computer Misuse Act, 1990).
- **8.15.3** The cases involved minors who were subjected to physiatrist test to determine their mental ability.
- **8.15.4** Admitted having committed the crime.

Therefore considering the decision of appeal court in (R v Mudd, 2017) to revise the sentence length from 24 months to 21 months persuaded the appeal court in (R v Connor Douglas Allsopp, 2019) to uphold the ruling of the junior court in line with the precedent set by the earlier case.

**8.16** On the issue of whether the case could have been suspended in lieu with the determination of blackmail case involving Daniel Kelly, accused of hacking into TalkTalk company, the court of appeal declined and sided with the Criminal court decision on (R v Connor Douglas Allsopp, 2019) that the delay of offending and trial should not be subjected to the two defendants. In my opinion however, two things arise;

- **8.16.1** Did the actions of appellant and Hanley lead to blackmailing the then TalkTalk CEO?
- **8.16.2** Were there any charges of blackmail preferred against the two together with Daniel Kelly and how will it affect the ruling/sentence?

In view of the court of appeal *horpeus corpus* case of (Zezev v Governor of Brixton Prison, 2002) where the co-accused had sought an appeal on extradition to United States of America following accusation of compromising systems and blackmailing Mr Bloomberg, the appeal court dismissed the plea whereby in the ruling Mr Justice Wright held in part that, "*Mr Zezev's at least apparent involvement, either in such conduct or in a conspiracy to be involved in such conduct, is plain on the evidence or* *discloses a prima facie case for him to answer under section 3 of Computer Misuse Act 1990."*

**8.17** Interpreting the arguments in the above case, am persuaded to mention that although the actions of the appellant and Hanley were linked to the subsequent attacks which lead to blackmail and demands for bitcoin, the court did not prefer blackmail charges against them, which to my opinion the extent of their involvement in the blackmail ought to have been propped and criminal culpability established based on the evidence adduced before the court. This therefore would have warranted suspension of the sentence until hearing and determination of Daniel Kelly case.

**8.18** In conclusion I submit that the junior court was more concerned with expedited justice rather than considering comprehensive aspects of laws contravened and the decision by appeal court to uphold the sentencing did not cure the problem either way. While I agree partly with the appeal court, considering the limitations to the issues raised, the utmost concern is driven by the fact that suspension of sentence plea was not issued as that would have established in details the criminal capability of the appellant in the subsequent attacks and blackmail of TalkTalk CEO.

**References**

ACPO, 2009. E-Crime strategy. *Association Of Chief Police Officers.*

Begovic, S., 2017. How to deal with E-crime. *E-bussiness development,* Volume 1, p. 5.

Casey, E., 2004. *Digital Evidence and Computer Crime.* 2nd ed. London: Elsevier Academic Press.

Computer Hope, 2019. *About Us: Computer Hope.* [Online]
Available at: https://www.computerhope.com/jargon/c/compcrim.htm
[Accessed 12 April 2019].

Computer Misuse Act, 1990. *C.18.* [Online]
Available at: https://www.legislation.gov.uk/ukpga/1990/18/contents
[Accessed 11 March 2019].

Criminal Damage Act , 1971. *C.48.* [Online]
Available at: http://www.legislation.gov.uk/ukpga/1971/48/contents
[Accessed 11 March 2019].

Data Protection Act, 1998. *C.29.* [Online]
Available at: http://www.legislation.gov.uk/ukpga/1998/29/contents
[Accessed 11 March 2019].

*Ellis v DPP* (2001)
https://login.westlaw.co.uk/maf/wluk/app/document?&srguid=i0ad8289e0000016981c00ae49d429d32&docguid=IA10F1A70E42711DA8FC2A0F0355337E9&hitguid=IA10F1A70E42711DA8FC2A0F0355337E9&rank=1&spos=1&epos=1&td=1&crumb-action=append&context=55&resolvein=true.

Fraud Act, 2006. *C.35.* [Online]
Available at: https://www.legislation.gov.uk/ukpga/2006/35/contents
[Accessed 11 March 2019].

Gercke, M., 2012. Phenomena, Challanges and Legal Response. *Understanding Cybercrime,* II(1), pp. 11-15.

Home affairs, 2013/2014. *UK Parliament.* [Online]
Available at: https://publications.parliament.uk/pa/cm201314/cmselect/cmhaff/70/70.pdf
[Accessed 11 March 2019].

laws.com, 2017. *cyber laws.* [Online]
Available at: https://cyber.laws.com/hacking
[Accessed 11 March 2019].

Malakedsuwan, P. &. S. K., 2019. A Model of E-fraud.

Microsoft, 2019. *Use of dump files.* [Online]
Available at: https://docs.microsoft.com/en-us/visualstudio/debugger/using-dump-files?view=vs-2019
[Accessed 9 April 2019].

Peda Net, 2019. *Computer Systems.* [Online]
Available at: https://peda.net/kenya/ass/subjects2/computer-studies/form-1/the-computer-system/nameless-89db
[Accessed 11 March 2019].

Protection to Harassment Act, 1997. *C.40.* [Online]
Available at: http://www.legislation.gov.uk/ukpga/1997/40/contents
[Accessed 11 March 2019].

*R v Connor Douglas Allsopp* (2019)
https://login.westlaw.co.uk/maf/wluk/app/document?&srguid=i0ad832f10000016981b69d086f5af7c0&docguid=I6D57DCC0351911E99376ECAA70B7D183&hitguid=I6D57DCC

0351911E99376ECAA70B7D183&rank=2&spos=2&epos=2&td=144&crumb-action=append&context=4&resolvein=true.

*R v Martin* (2013)
https://login.westlaw.co.uk/maf/wluk/app/document?&srguid=i0ad832f10000016981b69d086f5af7c0&docguid=I55F22AF0053011E39DB9E2BE65AF7566&hitguid=I55F22AF0053011E39DB9E2BE65AF7566&rank=44&spos=44&epos=44&td=144&crumb-action=append&context=11&resolvein=true.

*R v Mudd* (2017)
https://login.westlaw.co.uk/maf/wluk/app/document?&srguid=i0ad832f10000016981b69d086f5af7c0&docguid=IA7DD59A0A9AC11E7AE76A37EEA948475&hitguid=IA7DD59A0A9AC11E7AE76A37EEA948475&rank=14&spos=14&epos=14&td=144&crumb-action=append&context=4&resolvein=true.

Shon Harris, F. M., 2016. *Certified Information Systems Security Professional.* 7th ed. Indiana: McGraw-Hill education group.

Stefan Fafinski, W. H. D. &. H. M., 2010. *Mapping and Measuring Cybercrime.* Oxford, Oxford Internet Institute.

Stewart, 2006. *Collins Dictionary of law.* [Online]
Available at: https://legal-dictionary.thefreedictionary.com/electronic+crime
[Accessed 11 March 2019].

United Nations, 1994. United Nations. *Manual on the prevention and control of computer related crimes.*

Westlaw , 2019. *Westlaw.co.uk.* [Online]
Available at:
https://login.westlaw.co.uk/maf/wluk/app/document?&suppsrguid=i0ad6ada600000169722e14d868405fd4&docguid=IEFD73CB0353511E9900FAFBE922F562E&hitguid=I6D57DCC0351911E99376ECAA70B7D183&rank=2&spos=2&epos=2&td=144&crumb-action=append&context=4&resolvein=true
[Accessed 11 March 2019].

Yar, M., 2012. E-crime 2.0. *The Criminological Landscape of New Social Media,* 21(3), pp. 207-219.

*Zezev v Governor of Brixton Prison* (2002)
https://login.westlaw.co.uk/maf/wluk/app/document?&srguid=i0ad82d080000016981bdca139be24625&docguid=I122752D0E42911DA8FC2A0F0355337E9&hitguid=I122752D0E42911DA8FC2A0F0355337E9&rank=3&spos=3&epos=3&td=3&crumb-action=append&context=30&resolvein=true.